\documentclass{iau}
\usepackage{graphicx}
\usepackage{float}

\title[JD 11.~~Linking the central engine to the jet properties in radio loud AGN] 
{Linking the central engine to the jet properties in radio loud AGN}

\author[A.  Olgu\'in-Iglesias et al.]   
{A.  Olgu\'in-Iglesias$^1$, J. Le\'on-Tavares$^1$, V. Chavushyan$^1$, E. Valtaoja$^2$, C. A\~norve$^3$, K. Nilsson$^4$, J. Kotilainen$^4$, M. Tornikoski$^5$
}

\affiliation{$^1$Instituto Nacional de Astrof\'isica, \'Optica y Electr\'onica, Puebla, M\'exico,\\ $^2$Tuorla Observatory and Department of Physics, University of Turku, Finland,\\ $^3$Universidad Aut\'onoma de Sinaloa, M\'exico,\\ $^4$Finnish Centre for Astronomy with ESO, University of Turku, Finland,\\ $^5$Aalto University Mets\"ahovi Radio Observatory, Finland}

\pubyear{2014}
\volume{313}  
\pagerange{119--126}
\jname{Extragalactic Jets from every angle}
\editors{F. Massaro, C. C. Cheung, E. Lopez, A. Siemiginowska, eds.}
\begin{document}

\maketitle

\begin{abstract}
We explore the connection between the black hole mass and its relativistic jet  for a sample of radio-loud AGN ($z<1$), in which the relativistic jet parameters are well estimated by means of long term monitoring with the 14m Mets\"ahovi millimeter wave telescope and the Very Long Base-line Array (VLBA).  NIR host galaxy images taken with the NOTCam on the Nordic Optical Telescope (NOT) and retrieved from the 2MASS all-sky survey allowed us  to perform  a detailed surface brightness decomposition of the host galaxies in our sample and to estimate reliable black hole masses via their bulge luminosities. We present early results on the correlations between black hole mass and the relativistic jet parameters. Our preliminary results suggest that the more massive the black hole is, the faster and the more luminous jet it produces.

\end{abstract}

\firstsection 
\section{Introduction}
The relations between the black hole mass $(M_{BH})$ and its host galaxy bulge properties such as luminosity $L$, stellar velocity dispertion $\sigma*$ and mass $M$, (Gebhardt et al. 2000, Graham 2007, Kotilainen et al. 2007), suggest a coeval black hole-host galaxy evolution. It is widely believed that the relativistic jet launched by the black hole might play an important role in this co-evolution, by heating and forcing the cold gas out of the bulge, thus affecting star formation. So the question arises: Is there any relation between the mass of a black hole and the jet it produces? To effectively address this issue, we need reliable $M_{BH}$ estimations. Black hole masses can be estimated by assuming virialization of the clouds in the BLR. However, studies by Arshakian et al. 2010 and Le\'on-Tavares (2010, 2013) have shown that the gas in the BLR can be accelerated and ionized by non-thermal emission from the jet. Moreover, absorption features are swamped by non-thermal emission from the jet, thus making impossible to measure $\sigma*$. Finally, the empirical relation between the $M_{BH}$ and the bulge luminosity ($L_{bulge}$) can be used to obtain homogeneous and bias-free estimations of $M_{BH}$ in the overall population of AGN with prominent relativistic jets.

\section{$M_{BH}-jet$ relations}
\begin{figure}[H]
\centering
\includegraphics[scale=.53]{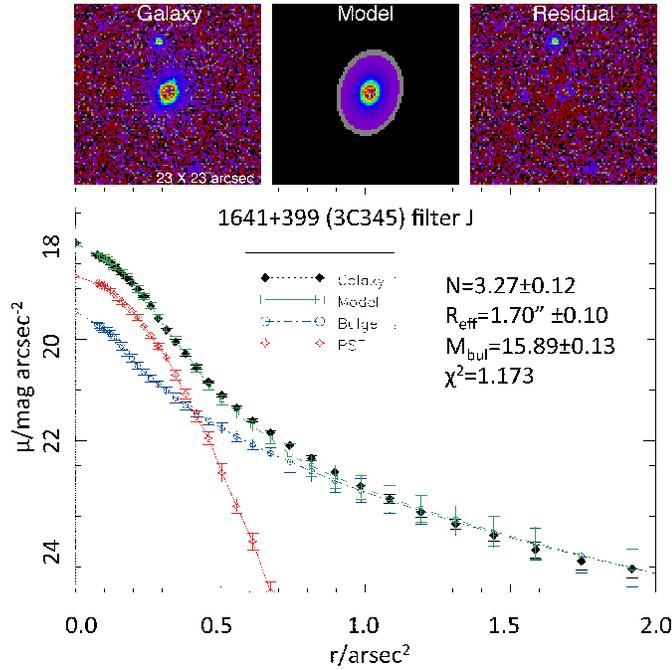}
 \caption{2D surface brightness decomposition of 1641+399 (3C345) at J band taken with with the NIR camera NOTCam on the Nordic Optical Telescope (NOT). Top left subpanel: the observed image. Middle top subpanel: Shows the model used to describe the surface brightness distribution, which is a nuclear unresolved component (PSF) convolved with a S\'ersic model. Right top subpanel: The residual image. Bottom: Radial profile of the surface brightness distribution.}
\label{profiles}
\end{figure}

By using the $M_{BH}-L_{bulge}$ relations (Graham 2007), we estimate the $M_{BH}$ of 29 radio loud AGN. Bulge luminosities were estimated by modeling the surface brightness of the host galaxy (figure \ref{profiles}) with GALFIT (Peng et al. 2011). Thus, allowing us to investigate the connection between $M_{BH}$ and jet properties such as luminosity (Torrealba et al. 2012) and Lorentz factor (Hovatta et al. 2009). We find statistically significant correlations between $M_{BH}$, jet luminosity and Lorentz factor.

\section{Implications}

Statistically significant correlations found between $M_{BH}$ and its jet in this work, strongly suggest that the more massive the black hole is, the faster and brighter the jet it produces. These result could be interpreted as an evolutionary connection between the black hole and its relativistic jet. Moreover, these relations could be useful to constraint high energy production models in radio-loud AGN. The manuscript reporting these findings in detail is under preparation.


\begin{thebibliography}{} 

\bibitem[Arshakian et al.(2010)]{arshakian_2010}{Arshakian, T.~G., 
Le{\'o}n-Tavares, J., Lobanov, A.~P., et al.\ 2010}, \textit{MNRAS}, 401, 1231 

\bibitem[Gebhardt et al.(2000)]{gebhardt_2000}{Gebhardt, K., Bender, 
R., Bower, G., et al.\ 2000}, \textit{ApJL}, 539, L13 

\bibitem[Graham(2007)]{graham_2007}{Graham, A.~W.}, 2007, \textit{MNRAS},379, 711


\bibitem[Hovatta et al.(2009)]{hovatta_2009}{Hovatta, T., Valtaoja, E., Tornikoski, M., {\ L\"a}hteenm{\"a}ki, A.}, 2009,\textit{AAP}, 494, 527

\bibitem[Kotilainen et al.(2007)]{kotilainen_2007}{Kotilainen, J.~K., 
Falomo, R., Labita, M., Treves, A., \& Uslenghi, M.}, 2007, \textit{ApJ}, 660, 1039

\bibitem[Le{\'o}n-Tavares et al.(2010)]{leon_tavares_2010}{Le{\'o}n-Tavares, J., Lobanov, A.~P., Chavushyan, V.~H., et al.} 2010, \textit{ApJ}, 715, 355 

\bibitem[Le{\'o}n-Tavares et al.(2013)]{leon_tavares_2013}{Le{\'o}n-Tavares, J., Chavushyan, V., Pati{\~n}o-{\'A}lvarez, V., et al.}2013, \textit{ApJL}, 763, L36


\bibitem[Peng et al.(2011)]{peng_2011}{Peng, C.~Y., Ho, L.~C., 
Impey, C.~D., \& Rix, H.-W.}2011, \textit{Astrophysics Source Code}, 4010



\bibitem[Torrealba et al.(2012)]{torrealba_2012}{Torrealba, J., 
Chavushyan, V., Cruz-Gonz{\'a}lez, I., et al.}, 2012, \textit{RMXAA}, 48, 9 





\end{thebibliography}
\end{document}